\documentclass[12pt, letterpaper]{article}
\usepackage{amsmath}
\usepackage{graphicx}
\usepackage{subcaption}
\usepackage[utf8]{inputenc}
\usepackage[parfill]{parskip}
\usepackage{booktabs}
\usepackage{etoolbox}
\patchcmd{\thebibliography}{\section*{\refname}}{}{}{}
\usepackage[left=3cm, right=3cm, top=2cm]{geometry}
\usepackage[affil-it]{authblk}
\usepackage{hyperref}

\makeatletter
\renewcommand\section{\@startsection{section}{1}{\z@}%
                                   {-3.5ex \@plus -1ex \@minus -.2ex}%
                                   {2.3ex \@plus.2ex}%
                                   {\normalfont\normalsize\bfseries\uppercase}}
\renewcommand\subsection{\@startsection{subsection}{2}{\z@}%
                                     {-3.25ex\@plus -1ex \@minus -.2ex}%
                                     {1.5ex \@plus .2ex}%
                                     {\normalfont\normalsize\bfseries}}%
\makeatother

\renewcommand\Authfont{\fontsize{12}{14.4}\selectfont}
\renewcommand\Affilfont{\fontsize{12}{10.8}\itshape}

\begin{document}

	\title{DaTscan SPECT Image Classification for Parkinson’s Disease}
	\author{Justin Quan \thanks{justin.quan@generationse.com}}
	\author{Lin Xu}
	\author{Rene Xu}
	\author{Tyrael Tong}
	\author{Jean Su \thanks{jean.su@generationse.com}}
	\affil{GenerationsE Software Solutions, Inc.}

	\date{\footnotesize \today}

	\maketitle

	\begin{abstract}
Parkinson’s Disease (PD) is a neurodegenerative disease that currently does not have a cure. In order to facilitate disease management and reduce the speed of symptom progression, early diagnosis is essential. The current clinical, diagnostic approach is to have radiologists perform human visual analysis of the degeneration of dopaminergic neurons in the substantia nigra region of the brain. Clinically, dopamine levels are monitored through observing dopamine transporter (DaT) activity. One method of DaT activity analysis is performed with the injection of an Iodine-123 fluoropropyl (\textsuperscript{123}I-FP-CIT) tracer combined with single photon emission computerized tomography (SPECT) imaging. The tracer illustrates the region of interest in the resulting DaTscan SPECT images. Human visual analysis is slow and vulnerable to subjectivity between radiologists, so the goal was to develop an introductory implementation of a deep convolutional neural network that can objectively and accurately classify DaTscan SPECT images as Parkinson’s Disease or normal.  This study illustrates the approach of using a deep convolutional neural network and evaluates its performance on DaTscan SPECT image classification.   
\par The data used in this study was obtained through a database provided by the Parkinson’s Progression Markers Initiative (PPMI).  The deep neural network in this study utilizes the InceptionV3 architecture, 1st runner up in the 2015 ImageNet Large Scale Visual Recognition Competition (ILSVRC), as a base model.  A custom, binary classifier block was added on top of this base.  In order to account for the small dataset size, a ten fold cross validation was implemented to evaluate the model's performance.
	\end{abstract}

	\newpage

	\section{Introduction}
Parkinson’s Disease (PD) is a neurodegenerative disease that is currently without a cure.  Early diagnosis of PD has been recognized to help with treatment, reducing the severity of symptoms, and providing an improved quality of life for those who suffer from PD \cite{Pagan2012, Murman2012}.  Clinical studies over the years have helped to develop our understanding of the etiology of PD.  Clinically, PD is characterized by the loss and degeneration of dopaminergic neurons in the substantia nigra brain region, resulting in the significant loss of dopamine production \cite{Zafar2019, Demaagd2015}.  Booth (2015) explains a widely used method to monitor dopaminergic neurons.  Dopamine transmission is regulated by dopamine transporters (DaT), which are densely populated within the substantia nigra region.  Analysis of DaT function through the use of an Iodine-123 fluoropropyl (\textsuperscript{123}I-FP-CIT) tracer and Single Photon Emission Tomography (SPECT) imaging in the substantia nigra, specifically the striatum, region of the brain (putamen and caudate structures), is currently among the recognized diagnosis tools for early PD detection \cite{Booth2015}.  The standard practice for DaT SPECT image classification requires human visual analysis performed by radiologists.  Unfortunately, this process is not only time consuming but also vulnerable to subjectivity and variability between observers.
\par Deep learning has become a fast growing field of machine learning.  The success and popularity of deep learning can be attributed to factors such as advancements in central processing units and graphics processing units, increased availability of large amounts of data, and developments of learning algorithms \cite{Shen2017}.  The popularity of using deep convolutional neural networks (CNN) for medical image classification has been on the rise and have led to some interesting fields of research \cite{Suk2015,  Heung2015, Dou2016, Cire2013}. 	
\par The motivation behind this study is to build upon the growing list of applications of deep learning in medical image analysis.  The goal of this study is to implement an introductory pipeline using a deep CNN that is able to efficiently, accurately and objectively classify DaTscan SPECT images as PD or non-PD.

	\section{Methods}
	\subsection{PPMI Data Collection}
Data used in this study was retrieved from a database provided by the Parkinson’s Progression Markers Initiative (PPMI).  PPMI is an observational clinical study to gather and verify progression markers in PD.  The aim of the PPMI study is to create a comprehensive set of clinical, imaging and biosample data that can be used to identify biomarkers of PD progression \cite{PPMI}. 
\par	The dataset that was collected for this study contains 659 unique patient DaT SPECT images.  These images belong to one of two classes: PD (n=449) and non-PD (n=210).  To ensure the model is trained on a set of unique subjects and tested on strictly unseen data, there are no images from follow-up patient visits within the dataset.

	\subsection{Image Preprocessing}
Raw \textsuperscript{123}I-FP-CIT SPECT images were acquired at PPMI imaging centers following their protocol \cite{PPMIprotocol}.  These raw images underwent reconstruction and preprocessing.  The data has been normalized and aligned to the Montreal Neurologic Institute (MNI) space, one of the accepted standard coordinate systems used for image registration.  Each DaT SPECT is finally presented as a 3D volume in the shape of 91 x 109 x 91.
\par With limitations in system memory capacity, this study uses the ImageDataGenerator class from the Keras deep learning library.  The ImageDataGenerator is able to load small batches of images on the fly, directly from their respective directories (train, validation and test) using the flow\_from\_directory method.  As the ImageDataGenerator cannot load volumetric Dicoms and currently does not support custom generators,  all the DaT SPECT images were converted into png format.  Each DaT SPECT image consists of a 3D volume of 91 slices.  After reviewing the methods of image preprocessing from other studies \cite{Zhang2016, Rumman2019, Prashanth2016} and visual analysis of each slice, it was decided that slices 40-42 appeared to illustrate the areas of interest with the highest pixel intensity. In order to keep each data image a unique subject and follow the 3-channel input format for the InceptionV3 model, the 3 slices were concatenated, creating a single 3-channel (RGB) image. Slice 40, 41, and 42 made up the RGB channels respectively.
\par To make use of the small dataset,  the images underwent augmentations via a series of transformations.  The model does not see the same image twice, which reduces the chance of the model overfitting on the training data.  Data augmentation allows deep learning to be performed with a small dataset by artificially increasing the size of the dataset through the creation of modified and transformed copies.  The augmentation ranges were small since SPECT scan images follow a protocol in order to prevent any discrepancies in alignment between scans.  Data augmentations were done on the fly, just prior to being progressively loaded to the network.  This is performed using the ImageDataGenerator class, provided by the Keras deep learning library.  Training data underwent width shifts, height shifts, horizontal flipping and brightness augmentations.

	\subsection{Dataset K-fold Split}
The dataset of 659 unique patient DaT SPECT images were split into 10 folds to be used for k-fold cross validation.  Each one of the images were randomly placed into one of the 10 folds while ensuring that the ratio of classes was kept the same for each fold.  As the dataset does not perfectly split into 10, the outcome was 9 folds consisting of 66 DaT SPECT images (21 control and 45 PD) and one fold of 65 DaT SPECT images (21 control and 44 PD).  Each unique fold was used once as a validation dataset, while the remaining 9 folds comprised the training dataset.  The split ratio of training and validation from the entire dataset was 9:1 respectively.  For 9 iterations, there were 593 training images and 66 validation images.  The 10th iteration consists of 594 training images and 65 validation images.  The mean of the validation loss and validation accuracies among the 10 iterations was used to evaluate performance.  The purpose for using 10-fold cross validation was to select high level hyperparameters before running a final model on test data.

	\subsection{Neural Network Architecture}
This study implements a deep CNN using the Keras neural network library that is running on top of a TensorFlow backend.  Since the dataset is on the smaller side (n=659), this study utilized transfer learning of a pre-trained model and defined a custom, binary classifying block. The pre-trained model used for transfer learning was InceptionV3 \cite{Szegedy2016}.  InceptionV3 was selected as it was one of the notable finalists in the 2015 ImageNet Large Scale Vision Recognition Competition (ILSVRC) \cite{ILSVRC}.  The InceptionV3 model has learned features from ImageNet’s large visual database \cite{Imagenet}.  The classifying block consists of a 2D global average pooling layer taking the base model’s output as its input, followed by a dense layer with ReLU activation, followed by a dropout layer, and a final dense layer with sigmoid activation.  This combined model was compiled using the Adam optimizer with an initial learning rate of $10^{-3}$ and a step decay function that gradually decreased the learning rate until it reached a final learning rate of $10^{-6}$.  The loss function used was binary crossentropy and the metric used to analyze the performance of the model was accuracy.  Each model of the ten-fold cross validation was trained over 500 epochs with a batch size of 16.

	\section{Results}
	\subsection{K-fold Cross Validation Performance Evaluation}
The K-fold cross validation performed 10 iterations of training and validating a model.  Each iteration used a unique fold as the validation dataset.  The validation loss and validation accuracy results for each iteration are presented in Table 1 below.  The test scores were then used to calculate a mean validation accuracy and loss.  Because the sample sizes of data was not consistent for one of the folds, the validation loss and accuracies were weighted when calculating the mean.  Although the results show that the training and validation accuracies are close, the validation loss is notably greater than the training loss among all the models.  This generally hints that the models may have been overfitting on the training data.  Iterations of adjustments to the dropout level were made, but the difference between loss values remain significant.  This suggests that the validation dataset size may be the cause.
	
	\ 

	\begin{table}[h]
	  \begin{center}
	    \begin{tabular}{c|c|c|c|c} 
	     \toprule
	      \textbf{Fold} & \textbf{Train Loss} & \textbf{Train Accuracy} & \textbf{Val Loss} & \textbf{Val Accuracy}\\
	      \hline
	      1 &  0.0088 & 1.0000 &  0.0848 & 0.9800\\
	      2 &  0.0108 & 1.0000 &  0.0459 & 0.9800\\
	      3 &  0.0968 & 0.9584 &  0.1729 & 0.9400\\
	      4 &  0.0148 & 0.9966 &  0.1395 & 0.9800\\
	      5 &  0.0603 & 0.9635 &  0.1180 & 0.9600\\
	      6 &  0.0341 & 0.9736 &  0.1591 & 0.9600\\
	      7 &  0.0033 & 1.0000 &  0.1178 & 0.9600\\
	      8 &  0.0033 & 1.0000 &  0.2117 & 0.9600\\
	      9 &  0.0049 & 1.0000 &  0.1985 & 0.9600\\
	      10 &  0.0029 & 1.0000&  0.0821 & 0.9796\\
	     \hline
	    \shortstack{\\Weighted\\Mean} & 0.0240 & \shortstack{0.9890\\($\pm$ 0.0161)} &  0.1329 &  \shortstack{0.9645\\($\pm$ 0.0128)}\\
	     \bottomrule
	    \end{tabular}
	    \caption{Ten fold cross validation results.}
	    \label{tab:table1}
	  \end{center}
	\end{table}

	\subsection{Final Model}
The final architecture and hyperparameters that were obtained from the 10-fold cross validation were used for the final model.  The original dataset of 659 DaT SPECT images were randomly split in a 80:20 ratio into a training dataset (n=527) and a testing dataset (n=132) respectively.  Both datasets were sorted into sub directories by their class and loaded into the model using the flow\_from\_directory() method from the ImageDataGenerator class.  The training and testing dataset maintain the same ratio of the two classes, 30\% control and 70\% PD.  The training images underwent the same augmentations and ranges as they were fed into the model.  The model was trained over 500 epochs with a batch size of 16.  The learning rate was initialized at $10^{-3}$ and ended at $10^{-6}$ through a gradual step decay.   
\par From the predictions of 132 test images, 88 were true positives, 1 was a false positive, 42 were true negatives, and 1 was a false negative.  This produced a sensitivity of 0.9888, a specificity of 0.9767 and a precision of 0.9888.  The results were plotted on a precision-recall (PR) curve and a receiver operating characteristic (ROC) curve.  The area under the PR curve was 0.9967 and the area under the ROC curve was 0.9909.   
	
	\ 

	\vspace{-0.25cm}

	\begin{table}[h!]
	  \begin{center}
	    \begin{tabular}{c|c|c|c} %
	     \toprule
	      \textbf{Test Accuracy} &  0.9848 & \textbf{PR auc} & 0.9967\\
	      \textbf{Test Loss} &  0.0656 & \textbf{ROC auc} & 0.9909\\
	      \hline
	      \textbf{True Positive} &  88 & \textbf{True Negative} & 42\\
	      \textbf{False Positive} &  1 & \textbf{False Negative} & 1\\
	      \textbf{Sensitivity} & 0.9888 & \textbf{Specificity} & 0.9767\\
	     \bottomrule
	    \end{tabular}
	    \caption{Performance results of final model.}
	    \label{tab:table3}
	    \ 
	  \end{center}
	\end{table}

	\newpage

	\par Performance measures were calculated using the following formulas:
	\begin{equation*}
		Sensitivity = \frac{True Positive}{True Positive + False Negative}
	\end{equation*}

	\begin{equation*}
		Specificity = \frac{True Negative}{True Negative + False Positive}
	\end{equation*}

	\begin{equation*}
		Precision = \frac{True Positive}{True Positive + False Positive}
	\end{equation*}

	\ 	

	\begin{figure}[h!]
	  \centering
	  \begin{subfigure}[b]{0.4\linewidth}
	    \includegraphics[width=\linewidth]{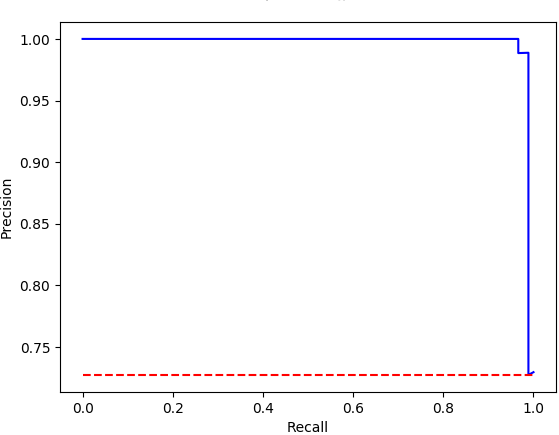}
	    \caption{PR curve}
	  \end{subfigure}
	  \begin{subfigure}[b]{0.4\linewidth}
	    \includegraphics[width=\linewidth]{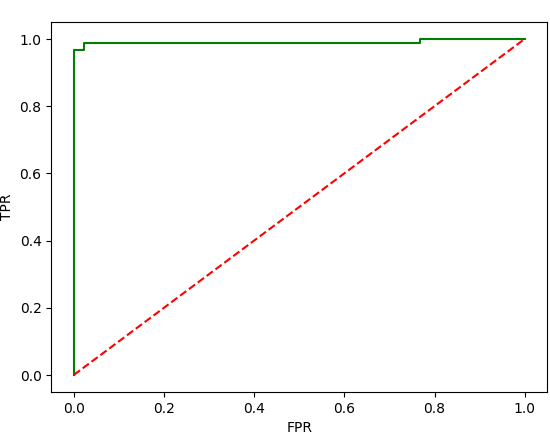}
	    \caption{ROC curve}
	  \end{subfigure}
	\caption{(a) PR curve with an area under the curve value of 0.9967.  (b) ROC curve with
			an area under the curve value of 0.9909.}

	\end{figure}

	\ 

	\vspace{-0.5cm}

	\begin{figure}[h!]
	  \centering
	  \begin{subfigure}[b]{0.3\linewidth}
	    \includegraphics[width=\linewidth]{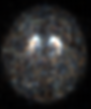}
	    \caption{False Positive}
	  \end{subfigure}
	  \begin{subfigure}[b]{0.3\linewidth}
	    \includegraphics[width=\linewidth]{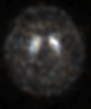}
	    \caption{False Negative}
	  \end{subfigure}
	\caption{Two incorrect predictions made by the final model.  Images presented are 3-channel (RGB) images with slices 40, 41, 42 
			concatenated.}
	\ 
	\end{figure}

	\newpage

	\section{Discussion}
There are a few limitations to this study.  The first being the small dataset size and imbalance of class data.  To maintain subject uniqueness, subsequent visit data were filtered out.  As neural network models are sensitive to data used in training, a small dataset means that each specific example will have a larger impact on the model’s classification skill.  A small testing dataset will likely lead to an optimistic, high variance estimation in the final model’s performance.  
\par The second limitation was the clinical, labeling of the DaTscan SPECT volumes used in this study.  As there currently isn’t a definitive diagnostic test for PD, neural network models require radiologists to accurately label the volumes as normal vs PD.  This means the data is still vulnerable to subjectivity of the human evaluations.
\par	The third limitation is that selective slices were used rather than the entire volumetric image.  This creates a selection bias by potentially ignoring important information found in the adjacent slices.  Due to limitations of the system memory capacity, images were  progressively loaded and augmented in small batches on the fly by using the ImageDataGenerator.
\par The fourth limitation of this study is that the dataset consists of subjects from a similar age group.  The model cannot be generalized as it has only been trained and tested on this specific age range.  It is also prevalent for individuals in this age range to have other mental health conditions which could affect the same region of interest and therefore affect the DaTscan SPECT images .
\par	Although there are still limitations, this study has shown that there is potential to develop a skillful model that can distinguish between normal and PD patients with little data.  From the results obtained, there is potential that deep neural networks can be used for accurate classification of Dicom images and that DaT SPECT images can be a viable input for training a machine learning model. 

	\section{Conclusion}
The purpose of this study was to implement an introductory pipeline that uses a deep convolutional neural network to classify DaTscan SPECT images in order to distinguish between normal subjects and those with Parkinson’s Disease.  Although the evaluated performance of this deep neural network model was favorable, with only one case of false positive and false negative respectively,  it is likely that these results were optimistically swayed due to the small dataset size and imbalance of class data.  Incorporating a larger dataset and modifications to input an entire volumetric image, the architecture of this study and its procedures can be useful with further research in this topic.

	\newpage
	\section*{References}
	\bibliography{bibliography}
	\bibliographystyle{ieeetr}

\end{document}